# Dependence of spin pumping spin Hall effect measurements on layer thicknesses and stacking order

V. Vlaminck[1,*], J. E. Pearson[1], S. D. Bader[1,2], and A. Hoffmann[1]

[1] *Material Science Division, Argonne National Laboratory, Argonne, Illinois 60439, USA*
[2] *Center for Nanoscale Materials, Argonne National Laboratory, Illinois 60439, USA*

Voltages generated from inverse spin Hall and anisotropic magneto-resistance effects via spin pumping in ferromagnetic (F)/non-magnetic (N) bilayers are investigated by means of a broadband ferromagnetic resonance approach. Varying the non-magnetic layer thickness enables the determination of the spin diffusion length in Pd of 5.5 ± 0.5 nm. We also observe a systematic change of the voltage lineshape when reversing the stacking order of the F/N bilayer, which is qualitatively consistent with expectations from spin Hall effects. However, even after independent calibration of the precession angle, systematic quantitative discrepancies in analyzing the data with spin Hall effects remain.

## I. INTRODUCTION

Spin Hall effects, which occur in non-magnetic metals due to spin-orbit coupling, give rise to the interconversion of spin and charge currents [1-4]. The transverse nature of these effects open interesting possibilities for magnetic memory and logic devices. This is due to the possibility of generating large net spin currents in thin film geometries [5,6], and connecting magnetization dynamics in ferromagnetic insulators with charge currents [7]. Recent theoretical [8-11] and experimental [5,6,12-20] work has identified relevant materials for spintronics applications based on spin Hall effects. Two experimental approaches have been developed to quantitatively investigate spin Hall effects: non-local injection in lateral spin valve structures [21-24], and electrically detected ferromagnetic resonance (FMR) in ferromagnetic (F)/non-magnetic metallic (N) bilayers [14,15,17,18], which is of interest in the present article. Large discrepancies exist among experiments to quantify the magnitude of the spin Hall angle, which describes the conversion efficiency between spin and charge currents [4]. Electrically detected FMR avoids complications of complex current flow patterns that appear in lateral spin valves that utilize metals with short spin diffusion lengths. However, the analysis of the FMR lineshape and, thus, the quantification of the spin Hall angle and conductivity remains unclear.

The difficulty comes from a superposition of different contributing mechanisms, such as anisotropic magneto-resistance (AMR), anomalous Hall effect (AHE), spin pumping-inverse spin Hall effect (SHE), and spin transfer torque (STT). Each can give rise to a voltage as a function of applied magnetic field. And the voltages can yield symmetric or anti-symmetric Lorentzian signatures depending on the measurement geometry (which includes the sample geometry, as well as applied field direction, and excitation field orientation). Another issue comes from the ability to unambiguously determine the relative phase between the electric *rf* currents in the sample and the magnetization dynamics, which may be both magnetic field and current driven [24]. In some cases this can obscure the electrically detected FMR lineshape. The issue of phase mixing is complex, as it appears to vary drastically with the sample geometry and frequency. Additional complications may arise from a non-uniformity of the microwave power along a coplanar waveguide [26], which can lead to systematic error in the analysis of the lineshape. Therefore, the choice of a suitable geometry that addresses the effect of interest is crucial.

We use a geometry that limits both the STT effect and the phase difference between the electric and magnetic dynamic field. We report a study of the electric detection of inverse spin Hall effects from spin pumping in permalloy (Py=$Ni_{80}Fe_{20}$)/N bilayers (N=Pd, Pt, Au) for different metal thicknesses and stack orders of the F/N bilayer. The paper is organized as follows. After discussing general experimental details, we present the metal thickness dependence of the lineshape and show how it enables the determination of the spin diffusion length of the metal. Next, we focus on the change of the lineshape when reversing the stacking order of the bilayers. Finally, we show how calibration of the precession cone angle resolves some inconsistencies in the experimental data.

## II. EXPERIMENTAL DETAILS

Figure 1(a) shows a schematic of the spin pumping - inverse SHE experiment. It consists of a 20-$\mu$m wide Py/Pt bilayers integrated with a 3-mm long, 30-$\mu$m wide and 150-nm thick coplanar wave-guide (CPW) transmission line made of Au. The bilayer, which is deposited directly on the substrate (undoped GaAs or intrinsic Si with a 300-nm thick layer of thermally grown $SiO_2$), is separated from the CPW by an 80-nm thick spacer layer of $SiO_2$ or MgO. Four electrical leads for the electrically detected FMR are connected to the Py/Pt bilayer, as shown in Fig. 1(b). In order to minimize inductively coupled currents in the sample it is essential to keep these electrical contacts between the central line and the ground plate within the CPW, since samples with contacts away from the CPW have shown complex variation of the phase of the electrically detected FMR with frequency. The sample is then wire bonded into a printed circuit board with a matching CPW and SMA connectors and mounted in an electromagnet, allowing a rotation in the plane of the sample. The relative angle between the transmission line and the applied field must be kept between 5 and 70°., (If it is 0°, the spin Hall voltage along the line vanishes, and if it is too large, the excitation field becomes parallel to the equilibrium direction and is inefficient.) The FMR is performed in transmission via either a vector network analyzer (Anritsu 37147C) or an amplitude modulation of the microwave source (Agilent 8257D) and lock-in detection using a diode. We work at constant frequency and sweep the magnetic field through the FMR. The frequency range of the measurement is kept between 3 and 11 GHz, as higher frequencies are no longer transmitted through the wire bonds. We also amplify the microwave power, while staying in the linear response regime, in order to measure a SHE signal in the tens of µV.

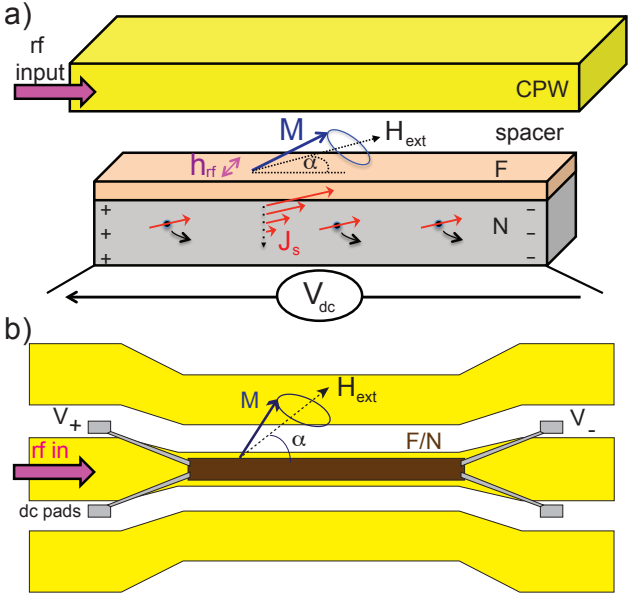

Figure 1: (a) Schematic of the spin pumping spin Hall effect experiment showing the respective polarity (*rf* input, external field and *dc* voltage contacts) used for the measurement. (b) Layout of voltage contacts used for the measurements (sketch not up to scale).

The *rf* current flowing in the wave-guide generates an *rf* magnetic field that excites magnetization dynamics in the permalloy layer. At resonance, the magnetization precession of the Py layer creates a spin accumulation at the F/N interface via the spin pumping mechanism [27]. The spin accumulation diffuses away from the interface, forming a gradient within the thickness of the N-layer. This leads, via the spin-orbit interaction, to a charge imbalance that is perpendicular to both the spin polarization $\boldsymbol{P}$ and the direction of the diffusion of the spin current [3]:

$$\boldsymbol{J_c} = e\gamma D (\boldsymbol{\nabla} \times \boldsymbol{P}) \qquad . \qquad (1)$$

Here $e$ is the electron charge, $D$ is the electron diffusion coefficient, and $\gamma$ is the spin Hall angle. Thus, the sign of the spin Hall voltage measured along the N-layer depends on the spin diffusion direction and, therefore, must change with the relative order of the bilayer stack. The spin Hall voltage is expected to have a symmetric Lorentzian lineshape, following the evolution of spin pumping via the FMR. This spin Hall voltage is measured separately via either lock-in amplification or with a *dc* voltmeter.

## III. THICKNESS DEPENDENCE

Figure 2(a) shows *dc* voltages measured at 7 GHz, for GaAs/Pd($t_{Pd}$)/Py bilayers with various thicknesses $t_{Pd}$ of Pd, while keeping the Py thickness constant at 15 nm. For a single layer of Py ($t_{Pd}$ = 0), where no SHE is expected, the measured voltage exhibits a purely anti-symmetric lineshape. As the thickness of the N-layer increases, the lineshape of the *dc* voltage acquires a symmetric component that saturates at $t_{Pd}$ =15 nm. Note that there is a difference in overall amplitude of the signal, which may originate from different *rf* transmission of the wire bonds for different samples [25]. Therefore, a direct comparison of the voltage amplitude between each sample is not straightforward.

The anti-symmetric component of the single layer linewidth is attributed to a homodyne AMR effect [28,29], which originates from an *rf* electrical current generated in the sample from either capacitive coupling to *rf* electric fields in the waveguide or inductive coupling to *rf* magnetic fields. The magnitude of this homodyne AMR voltage is given by [14,15]:

$$V_{AMR} = I_{FN} \Delta R_{AMR} \frac{\sin 2\theta}{2} \frac{\sin 2\alpha}{2} \cos\phi_0 \qquad . \qquad (2)$$

$I_{FN}$ corresponds to the part of the microwave current flowing in the bilayer. To a first approximation, $I_{FN}$ is assumed proportional to the current in the CPW, according to the ratio of the CPW and the bilayer *dc* resistance. $\Delta R_{AMR}$ is the bilayer absolute *dc* magneto-resistance when changing the magnetization by 90° with respect to the current direction. The angles θ, α and $\phi_0$ correspond to the largest value of the precession cone angle, the angle of the applied field with respect to the central line, and the phase difference between the microwave field and the magnetization dynamics, respectively.

The geometry of our sample design, where the voltage contacts are close to the signal line of the CPW, minimizes inductive coupling and results mainly in *rf* currents in the sample due to capacitive coupling. In other words, the *rf* currents in the sample and the CPW are in phase with each other, resulting in an anti-symmetric Lorentzian voltage signal due to the continuously varying phase $\phi_0$ of the magnetization dynamic response with respect to the excitation field (*i.e.*, in-phase far below resonance, π/2 at resonance, and out-of-phase far above resonance).

In contrast, the inverse SHE contribution is independent of the phase of the magnetization dynamics and gives rise to a purely symmetric Lorentzian voltage contribution given by [14,15]:

$$V_{SHE} = \frac{\gamma e L E f\, g_{mix}\, \lambda_s\, \sin\alpha\, \sin^2\theta}{\frac{t_N}{\rho_N} + \frac{t_F}{\rho_F}} \tanh\left(\frac{t_N}{2\lambda_s}\right) \qquad . \qquad (3),$$

where, $L$, $E$, $f$, $\lambda_s$ and $\rho_N$ ($\rho_F$) correspond to, the length of the bilayer, a correction factor for the ellipticity of the magnetization precession [29], the frequency, the spin diffusion length in the normal metal, and the normal metal (ferromagnet) resistivity. The effective spin mixing conductance $g_{mix}$ determines the flow rate of spin pumped into the metal, taking into account the backflow of spin into the ferromagnet. The term $\tanh(t_N/2\lambda_s)$ describes the decay of the spin accumulation in the metal away from the F/N interface.

In the linear response regime, both the inverse SHE and the AMR components have the same power dependence (proportional to the square of the microwave field), so that the resultant $dc$ voltage is a sum of a symmetric ($V_S$) and an anti-symmetric ($V_A$) Lorentzian: $V_{dc} = W V_A + (1-W) V_S$, where $W$ represents the weight of the anti-symmetric component, When $W=1$ ($W=0$), the signal lineshape is fully anti-symmetric (symmetric).

Figure 2(b) summarizes the metal thickness dependence of the $dc$ voltage lineshape measured for $\alpha = 40°$, by plotting the weight $W$ versus Pd thickness. Only four frequencies (4, 6, 8 and 10 GHz) are shown for the sake of clarity. Starting from a single layer of Py ($t_{Pd} = 0$), the lineshape of the $dc$ voltage remains anti-symmetric ($W=1$) throughout the frequency range. This shows, in particular, that for a single ferromagnetic layer, no additional phase difference occurs between the $rf$ current and the $rf$ field, unlike what was suggested recently [17,25]. As the thickness of Pd increases, the weight $W$ decreases until it reaches a minimum value ($t_{Pd}\sim 15$ nm) where it remains mostly constant. Note that this behavior is qualitatively similar independent of the measurement frequency. For $t_{Pd} > \lambda_s$, the spin diffusion length, no SHE occurs above $\lambda_s$ as the spin accumulation has fully relaxed; therefore W reaches a plateau. The decrease of the weight with the Pd thickness appears more pronounced at higher frequency. This observation is related to the ellipticity of the magnetization precession, which reduces the spin pumping effect, and thus, lowers the relative amplitude of inverse SHE vs. AMR at lower frequency [30].

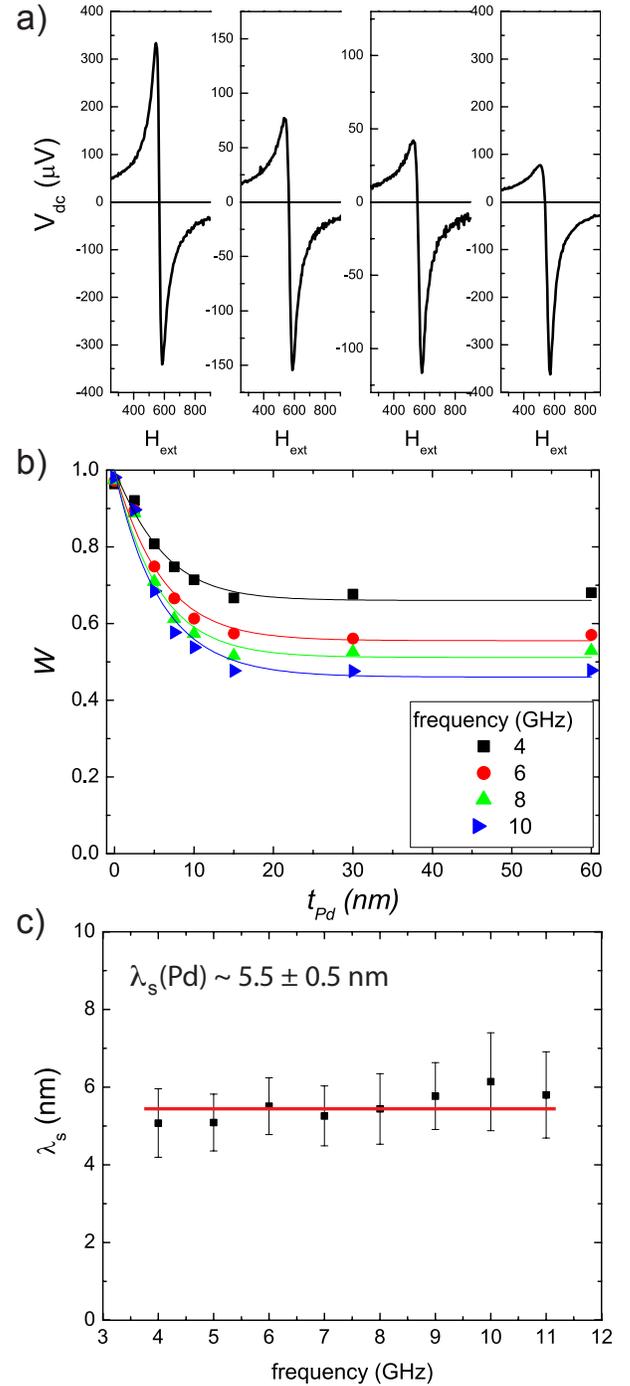

Figure 2: (a) AMR-SHE spectra for (GaAs/xPd/15Py) ($x$ = 0, 5, 10, 15 nm) measured at 7 GHz. (b) Thickness dependence of the weight $W$ of the anti-symmetric component for Pd. (c) Estimation of the spin diffusion length for Pd for all measured frequency.

Using Eq. 2, the precession angle $\theta$ can be directly estimated from the AMR voltage contribution assuming that the precession angle is uniform along the whole length and that the correlation between the $rf$ field $h_{rf}$ from the CPW and the microwave current flowing in the bilayer is well established. However, it was recently shown that these assumptions do not hold [26], and, as will be shown, these assumptions can lead to large discrepancies in the estimates of the spin Hall angles when reversing the bilayer stacking order. In the limit of small precession angle $\theta = h_{rf} \cos(\alpha)/\Delta H$ ($\Delta H$ being the half linewidth at half

maximum), the ratio of the maxima, $V_{SHE}/V_{AMR}$, can be written as:

$$\frac{V_{SHE}}{V_{AMR}} = \frac{\gamma e E f g_{mix} \lambda_s}{R_{CPW} I_{CPW} \frac{\Delta R_F}{R_F}} \frac{h_{rf}}{\Delta H} \frac{\rho_F L}{t_F} \tanh\left(\frac{t_N}{2\lambda_s}\right), \quad (4)$$

where $R_{CPW}$ and $I_{CPW}$ correspond to the resistance and microwave current in the CPW, and $\Delta R_F/R_F$ is the relative change of magneto-resistance for the ferromagnet (typically ~0.01 for Py). Equation 4 shows that the metal thickness dependence of $V_{SHE}/V_{AMR}$ is only contained in the distribution of the spin accumulation $\tanh(t_N/2\lambda_s)$, and that above a certain thickness ($t_N > \sim 3\lambda_s$), the ratio $V_{SHE}/V_{AMR}$ becomes independent of the metal thickness. In order to fit the thickness dependence of the weight $W$, which is related to $V_{SHE}/V_{AMR}$ by the relation $W=1/(1+|V_{SHE}/V_{AMR}|)$, we use a function of the form $1/(1+A \tanh(t_N/2\lambda_s))$, where $A$ is a constant independent of the metal thickness. In this way, we can directly extract $\lambda_s$ from the fitting, and verify, as shown in Fig. 2(c), that $\lambda_s$ remains constant for all measured frequencies. This alternative technique yields a value of $\lambda_s = 5.5\pm0.5$ nm for Pd, which is somewhat smaller than previously reported values measured at lower temperatures [16,31] and comparable to other values measured for Pd at room temperature [32,33].

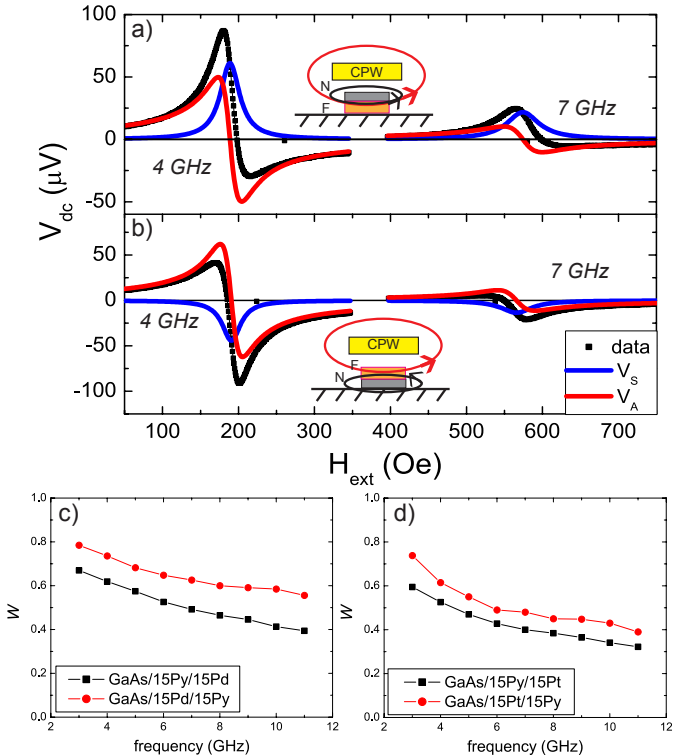

Figure 3: (a) AMR-SHE spectra for a GaAs/15Py/15Pd sample at 4 and 7 GHz, 40° applied field angle and 5 dBm. (b) Measurements under identical conditions for the reverse stack order (GaAs/15Pd/15Py). Frequency dependence of the weight $W$ of the anti-symmetric component of the AMR-SHE voltage for both stacking orders with Pd (c) and Pt (d).

## IV. STACKING ORDER DEPENDENCE

We now look at the change of the *dc* voltage upon reversing the stacking order of the F/N bilayer. Figures 3(a) and (b) show *dc* voltages measured for GaAs/15Py/15Pd and GaAs/15Pd/15Py bilayers respectively, at 4 and 7 GHz, 40° applied field angle, and 5 dBm *rf* power. Starting from the F/N configuration shown in Fig. 3(a) with the same polarity as in Fig. 1(a) for the *dc* contacts, the input port of microwave, and the direction of the applied field, both the anti-symmetric ($V_A$) and symmetric ($V_S$) components of the voltage are positive. We keep the same measurement polarity for the reverse configuration (N/F) in Fig. 3(b) and observe that only the symmetric component has changed sign. The anti-symmetric component, which is understood as a heterodyne AMR effect, keeps the same sign in both configurations, as expected, since it only depends on the relative polarity of the microwave current, the *dc* contacts, and the field orientation, which are identical in both measurements. The change of sign of the symmetric component is consistent with the cross-product relation of the inverse SHE [see Eq. (1)]. In the F/N configuration, where the Py is underneath the metal, the pure spin current pumped into the Pt diffuses upward and, according to Eq. (1), gives rise to a positive inverse spin Hall voltage, whereas in the N/F configuration the pure spin current diffuses downward, and the resultant SHE voltage is negative.

However, in addition to the lineshape change given by the sign reversal of the symmetric component, we also notice that the relative amplitudes of the symmetric and anti-symmetric components are not identical. The amplitude of $V_S$ with respect to $V_A$ in the F/N configuration appears larger than in the N/F configuration. This change of the lineshape with the stacking order is summarized in Fig. 3(c), by showing the weight of the anti-symmetric amplitude $W$ for the whole frequency range. The value of $W$ remains higher in the N/F than in the F/N configuration over the whole frequency range, meaning that the lineshape is more anti-symmetric for N/F *vs*. F/N. The difference in $W$ between F/N and N/F, which is ~20% at 3 GHz, increases at higher frequency, where it is ~30%. We also observed similar trends with Pt, which is also shown in Fig. 3(c), and where we used the same thicknesses (15 nm Py, 15 nm Pt, and 150 nm of Au for the CPW). The value of $W$ in the case of Pt appears slightly lower than in the case of Pd (meaning that the relative amplitude of the symmetric component is larger in the case of Pt). This is expected, as the spin Hall angle is known to be larger for Pt than for Pd [14-16]. However, the difference between the two stacking orders is similar for Pd.

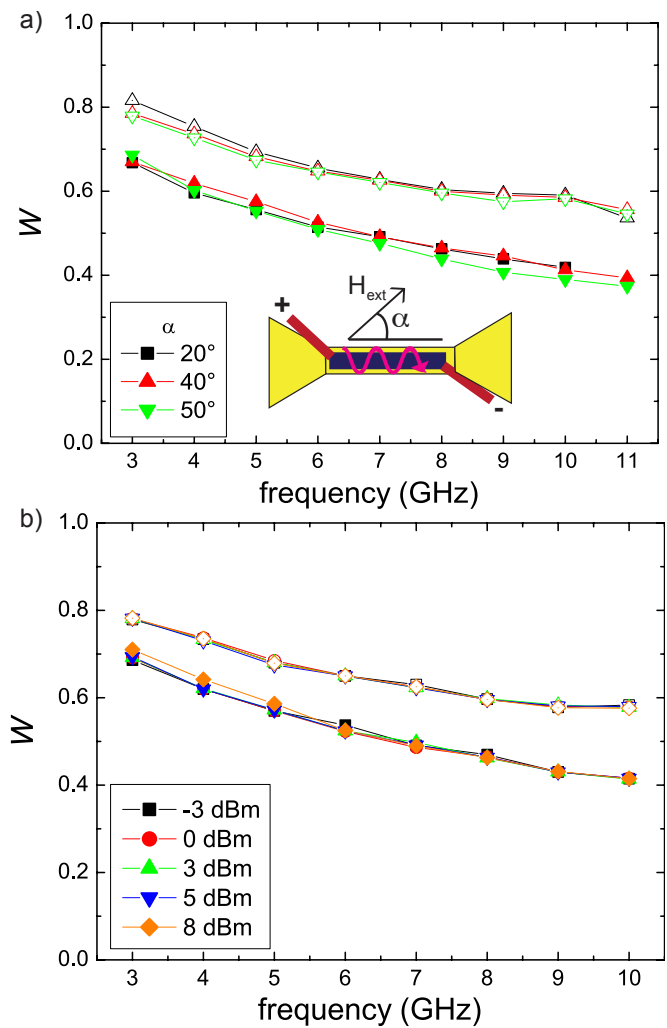

Figure 4: Frequency dependence of *W* for a GaAs/15Py/15Pd (solid symbols) and a GaAs/15Pd/15Py (open symbols). The lineshape remains unchanged with applied field angle (a) and *rf* output power (b).

To further investigate the change in the lineshape when changing the stacking order of the F/N bilayer, we verify in Fig. 4 that the lineshape remains unchanged with applied field angle and microwave power, as is expected from the expression of $V_{SHE}/V_{AMR}$ (see Eq. 4). We also measured identical devices on a different substrate Si/SiO$_2$ (not shown) and found the same trend, $W(N/F) > W(F/N)$, with a slightly reduced difference between the two configurations. Along with these changes in the lineshape between the two configurations, which have been observed systematically on several different samples, we also observed a systematic reduction of the absolute *dc* AMR as well as $M_s$ in the F/N configuration [see inset Figs. 5(a) and (b)]. However, this reduction of the *dc* AMR, which may be due to a difference in the growth of the Py, whether it is deposited first on the substrate or on the metal layer, does not account for the difference of a factor of two, when estimating the spin Hall angle from the two configurations.

A possible origin of this difference in the lineshape could be the additional Oersted field coming from rf currents passing through the metal layer. In the F/N configuration, where both the CPW and the metal sit on top of the Py layer, the two *rf* fields $h_{CPW}$ and $h_N$ add up, whereas in the N/F configuration, where the Py layer is in between the CPW and the metal layer, the two fields oppose each other. Therefore, according to the expression of $V_{SHE}/V_{AMR}$ (Eq. 4), the ratio of SHE and AMR amplitude at a given microwave power ($I_{CPW}$) ought to be larger in the F/N configuration where the net resulting *rf* field that the Py sees is larger than in the N/F configuration. As it was shown recently, this change of the microwave field in the Py layer between the two stacking orders could also be related to a difference in microwave screening from eddy currents [34], although the thicknesses of the layers in our samples are much smaller than the skin depth. It appears, therefore, essential to obtain an independent calibration for the microwave fields in the Py.

## V. CALIBRATION OF PRECESSION ANGLE

We adopt the AMR-FMR method of Costache *et al.* [35] to determine the precession angle in our samples independent of assumptions about the *rf* magnetic fields. The method consists in measuring the reduction of the AMR due to the opening angle of the magnetization at resonance. Similar to the SHE-AMR measurement, we excite the FMR via an amplitude modulation of the microwave power and measure the AMR-FMR with lock-in amplification. The carrier frequency of the modulation, which we vary between 1-50 kHz, does not matter for the measurement. We apply the magnetic field along the central line ($\alpha=0$), where neither SHE nor heterodyne AMR voltages occur, and apply a constant *dc* current ($I_{dc}$ = 10 to 20 mA) throughout the entire measurement. Prior to the measurement, we allow at least 10 min. for the system (sample + circuit board) to reach thermal equilibrium. Figures 5(a) and (b) show AMR-FMR measurements for SiO$_2$/15Py/15Pd and SiO$_2$/15Pd/15Py samples, respectively, at 6 GHz for three output powers (5, 8 and 11 dBm). The AMR-FMR measurements, which are sensitive to a voltage change due to modulation of the magnetization dynamics, have a Lorentzian lineshape at resonance. Away from the FMR, the absence of magnetization dynamics results in no voltage change. As the field approaches the resonant condition, the angle between the magnetization and the direction of the current starts to increase as the precession angle opens up, resulting in a change of the lock-in voltage equal to: $\Delta V = \Delta R_{AMR} I_{dc} \sin^2\theta$, where $\Delta R_{AMR}$ is the maximum change of AMR at 90° angle. The values of the precession angle at resonance $\theta_{res}$ follow an increase of $\sqrt{2}$ as the power is doubled ($\theta_{res} = h_{rf}/\Delta H$ and $h_{rf} \sim \sqrt{P_{rf}}$), which confirms the linear response regime of our measurement.

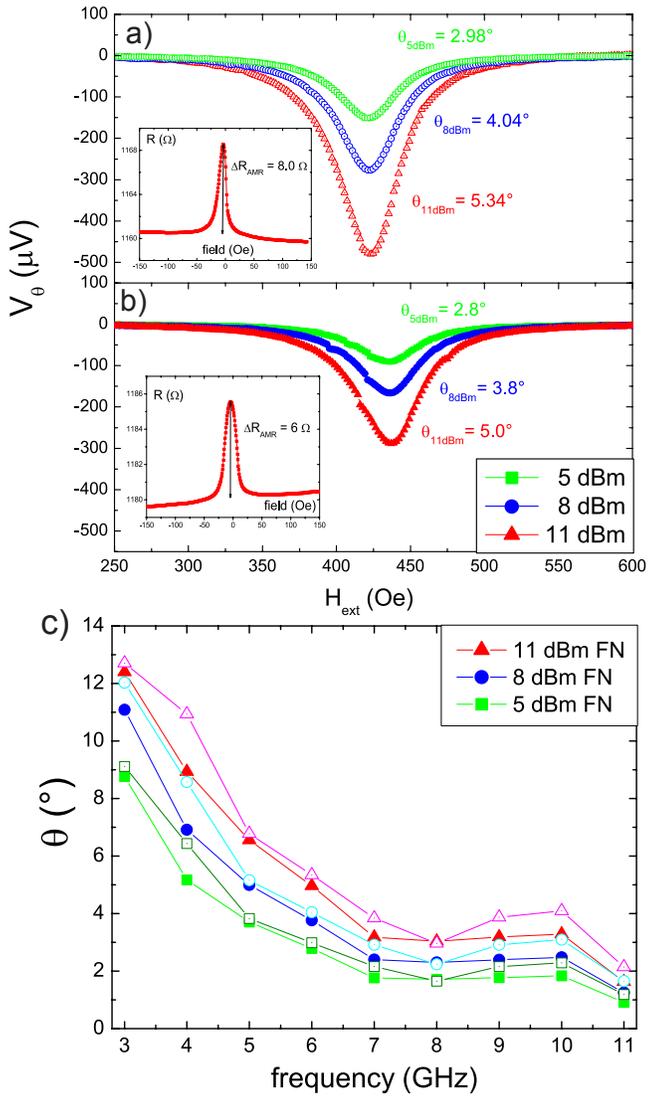

result in varying standing-wave patterns for the microwaves in the samples [26].

Considering possible spatial inhomogeneities of the microwave field along the sample, calibrating the cone angle with the AMR-FMR method has the advantage that its dependence on the cone angle is identical to that obtained from the SHE voltage (see Eq. 3), namely both are proportional to $\sin^2\theta$. In other words, an average value of $\sin^2\theta$ determined via the AMR-FMR measurement contains any non-uniformities of the *rf* field along the line and also across the wire due to the skin effect.

Finally, only the amplitude of the symmetric component of the SHE-AMR signal is needed for the estimation of the spin Hall angle γ, as we can directly implement the determination of $\sin^2\theta$ via the above calibration. All other parameters of Eq. 3 are known except the spin-mixing conductance, whose estimation remains a topic of debate due to complications, such as proximity effects and spin backflow at the interface [33,36,37]. We use $\lambda_s$(Pd) = 5.5 nm for Pd, as found from the thickness dependence of the lineshape).

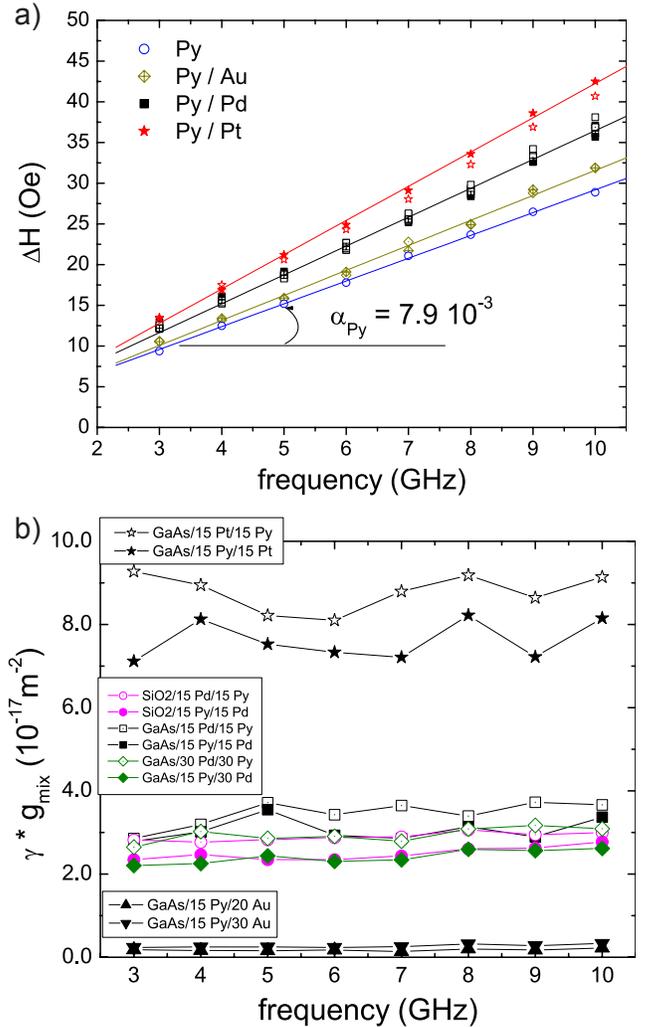

Figure 5: AMR-FMR measurements at 6 GHz and *rf* output power 5, 8, and 11 dBm for (a) SiO$_2$/15Pd/15Py (open symbols) and (b) SiO$_2$/15Py/15Pd (solid symbols). Insets in (a) and (b) show the corresponding *dc* magneto-resistance with the field applied perpendicular to the long direction (easy axis) of the samples. (c) Frequency dependence of the precession angle at resonance.

Note that the amplitude of *ΔV* differs between the F/N [Fig. 5(a)] and N/F [Fig. 5(b)] configurations, although both were measured under identical conditions of microwave power and *dc* current. Part of the difference in the amplitude of *ΔV* comes from a reduction of the AMR effect in the F/N configuration, as shown in the insets of Figs. 5(a) and (b). Furthermore, the difference in the value of θ$_{res}$ between the two configurations also illustrates the difference of microwave transmission. As the samples are connected via wire bonds to the circuit board, the microwave transmission is not only different from sample to sample, but is also frequency dependent and drops continuously throughout the frequency range, especially beyond 11 GHz. The frequency dependence of θ$_{res}$, as shown in Fig. 5(c), indicates that the SiO$_2$/15Pd/15Py sample has a minimum microwave transmission at 8 GHz, whereas the power transmission remains fairly constant between 7 and 10 GHz for the SiO$_2$/15Py/15Pd sample. This different behavior may reflect differences of impedance mismatch for different samples, which may

Figure 6: (a) Frequency dependence of the FMR linewidth showing the additional damping due to spin pumping. (b) Estimation of the product spin mixing conductance * spin Hall angle for Pt, Pd and Au following the independent calibration of the cone angle.

Although we did not study the thickness dependence of the lineshape for Pt or Au, we used $\lambda_s(Pt) = 4$ nm and $\lambda_s(Au) = 35$ nm as reported in the literature [16,31,38] and assuming that $\lambda_s(Pt) < \lambda_s(Pd)$. However, we note that the actual value of the spin diffusion length in Pt is controversial, [39, 40] with reported values ranging from 1-10 nm. Nevertheless, the quantitative analysis of the spin Hall angle based on Eq. 3 is only weakly dependent on the actual value of the spin diffusion length, as discussed in Refs [4, 15].

The resistivities measured for each sample have a thickness dependence ranging from 42 to 17 μΩcm between 2.5 and 60 nm for Pd, and are reproducible for identical thicknesses. Other values for our films are $\rho(Pt,15nm)=25$ μΩcm, $\rho(Au, 30$ nm$)=5$ μΩcm, and $\rho(Py, 15nm)=40$ μΩcm. The frequency dependence of the FMR linewidth presented in Fig. 6(a) shows additional damping due to spin pumping for all samples. One can see that the slope of this frequency dependence of the FMR linewidth, which is proportional to the total magnetic damping, is greater for Pt than for Pd, which is also greater than for Au. This increase of the additional damping between Au, Pd and Pt is in accordance with the spin pumping effect and spin back flow theory that was recently established [39]. From the measurement of this additional damping, we estimate the effective spin mixing conductance $g_{mix}$ that takes into account the spin backflow for each metal following the formalism of Tserkovniak *et al.* [27]. As can be observed in Fig. 6(a), we estimate that the spin mixing conductance from the broadening of the linewidth still has ~20% uncertainty, as additional effects, such as magnetic proximity [36], could also contribute to the FMR linewidth. Therefore, we summarize in Fig. 6(b), for each frequency and stacking configuration for Pd, Pt and Au, the measured products $g_{mix}*\gamma$, for which, according to Eqn.(3), we have better accuracy. We also report in Table (1) the spin Hall angles for each metal for the corresponding value of the spin mixing conductances. The values are $\gamma=0.012\pm0.002$ and $g_{mix} = 2.3 \pm 0.4 \cdot 10^{19}$ m$^{-2}$ for Pd, $\gamma=0.027\pm0.005$ and $g_{mix} = 3.0 \pm 0.6 \cdot 10^{19}$ m$^{-2}$ for Pt, and $\gamma=0.0025\pm0.0008$ and $g_{mix} = 0.9 \pm 0.2 \cdot 10^{19}$ m$^{-2}$ for Au.

| Metal | $\rho$ [μΩ.cm] | $\lambda_s$ [nm] | $\alpha$ [10$^{-2}$] | $g_{mix}$ [10$^{19}$m$^{-2}$] | $\gamma$ (%) |
|---|---|---|---|---|---|
| Py | 40 | 5 | 0.79 ±0.05 | | |
| Au | 5 | 35 | 0.91 ±0.18 | 0.9 ±0.2 | 0.25 ±0.08 |
| Pd | 20 | 5.5 | 1.09 ±0.22 | 2.3 ±0.4 | 1.2 ±0.25 |
| Pt | 25 | 4 | 1.18 ±0.24 | 3.0 ±0.6 | 2.7 ±0.5 |

Table 1: Spin mixing conductance and spin Hall angle reported for the various metal/Py samples measured, with their corresponding resistivity and spin diffusion length.

The AMR-FMR calibration of the cone angle reduced the difference in the $\gamma$–estimates for the two stacking order configurations. However, our refined analysis systematically shows a 20% larger $\gamma$ value in the N/F configuration for Pd and Pt. This remaining difference in the estimate of $\gamma$ is no longer related to a change of *rf* field, as this problem was addressed by the independent calibration of sin$^2\theta$. Another possible way to account for this difference could be related to a phase offset of the magnetization dynamics dependent of the stacking order, as was recently demonstrated by Bailey *et al.* via x-ray magnetic circular dichroism [40]. This change of phase could directly affect the asymmetry of the SHE-AMR lineshape.

## Conclusion

We have demonstrated that the symmetric component of the SHE-AMR lineshape initially increases with normal metal thickness and then becomes thickness independent. The lineshape change with metal thickness is explained within a previously used framework that takes into account both the homodyne AMR and the inverse SHE. This enables direct determination of the spin diffusion length of the normal metal. Furthermore, we showed that the symmetric component of the SHE-AMR voltage changes sign when the stacking order of the bilayer is reversed, in agreement with the cross product relation of the inverse SHE. In addition, there is a systematic change of the SHE-AMR lineshape upon reversal of the normal metal/ferromagnet stacking order. The relative amplitude of the symmetric component was always found to be larger in the case of the substrate/F/N configuration. To a large extent, this systematic difference in the lineshape between F/N and N/F can be understood as a difference in the *rf* magnetic excitation field $\boldsymbol{h}_{rf}$ in the permalloy. This can be accounted for by using an approach based on the change of magneto-resistance at the ferromagnetic resonance that enables the independent estimate of the precession cone angle for each frequency. The final estimate of the spin Hall angle using this independent calibration of the precession angle is therefore improved.

## Acknowledgement
We would like to thank F. Y. Fradin, R. Winkler and H. Schultheiss for illuminating discussions. This work and the use of the Center for Nanoscale Materials at Argonne National Laboratory are supported by the U.S. Department of Energy, Office of Science, Basic Energy Science under contract No. DE-AC02-06CH11357.